\documentclass{elsart}
\usepackage{graphicx}
\begin{document}
\begin{frontmatter}
\journal{Surface Science}
\title{Interaction between electronic structure and strain in Bi
  nanolines on Si(001)}
\author[NRI,NML]{J.H.G.~Owen\corauthref{jhgo}\thanksref{where}},
\ead{james.owen@materials.ox.ac.uk}
\corauth[jhgo]{Corresponding author}
\author[NRI,NML]{K.Miki} and 
\ead{miki.kazushi@aist.go.jp}
\author[UCL]{D.R.~Bowler\thanksref{LCN}}
\ead{david.bowler@ucl.ac.uk}
\ead[url]{http://www.cmmp.ucl.ac.uk/$\sim$drb/research.html}
\address[NRI]{Nanotechnology Research Institute (NRI),
National Institute of Advanced Industrial Science and Technology (AIST),
1-1-1 Higashi, Tsukuba. Ibaraki 305-8562, Japan}
\address[NML]{Nanomaterials Laboratory (NML), 
National Research Institute for Materials Science (NIMS)
Sengen 1-2-1, Tsukuba. Ibaraki 305-0047, Japan,}
\address[UCL]{Department of Physics and Astronomy, University College London,
Gower Street, London WC1E 6BT, UK}
\thanks[where]{Present address : Dept. of Materials, 
Oxford University, Parks Rd,Oxford, OX1 3PH, UK}
\thanks[LCN]{Also at: London Centre for Nanotechnology, Department of Physics and Astronomy, Gower Street, 
London WC1E 6BT, UK}

\begin{abstract} 
Heteroepitaxial strain can be a controlling factor in the lateral
dimensions of 1-D nanostructures. Bi nanolines on Si(001) have an
atomic structure which involves a large sub-surface reconstruction,
resulting in a strong elastic coupling to the surrounding silicon.  We
present variable-bias STM and first principles electronic structure
calculations of the Bi nanolines, which investigates this interaction.
We show that the strain associated with the nanolines affects the
atomic and electronic structure of at least two neighbouring Si
dimers, and identify the mechanism behind this.  We also present
partial charge densities (projected by energy) for the nanoline with
clean and hydrogenated surroundings and contrast it to the clean
Si(001) surface.
\end{abstract} 

\end{frontmatter} 

\section{\label{sec:intro}Introduction} 
Nanometer-scale electronic technologies require not only the formation
of nanoscale devices, but also nanoscale interconnections between the
devices. At present, nanoscale structures may be made via at least
three different approaches, all of which are likely to used in any
scheme: top-down lithographic methods, such as e-beam lithography and
AFM lithography; bottom-up methods, which involve positioning of
\textit{ex situ} ``prefabricated'' structures such as conducting
molecules, semiconductor nanowires\cite{Huang2001} and carbon
nanotubes\cite{Bachtold2001} using, e.g., scanning probes or
micro-fluidics; or by \textit{in situ} self-assembly, as in
semiconductor quantum dots, or silicide nanowires\cite{Chen2000}. We
are pursuing a hybrid self-assembly-based fabrication route, whereby a
Bi nanoline\cite{Miki1999a,Miki1999b} is used as a nanowire or as a
template for nanowires of other metals.  These nanolines are very
long---over 400 nm in some cases---and straight: a kink in a nanoline
has never been seen. Unlike silicide nanowires\cite{Chen2000}, their
width is constant, occupying the space of 4 dimers (1.5 nm) in the
Si(001) surface. However, variable-bias STM of the nanolines indicates
that they have a band gap which is \textit{larger} than the
surrounding surface, and suggests that they are not
conductive\cite{Miki1999a,Miki1999b}.

\begin{figure} 
\includegraphics[width=\textwidth]{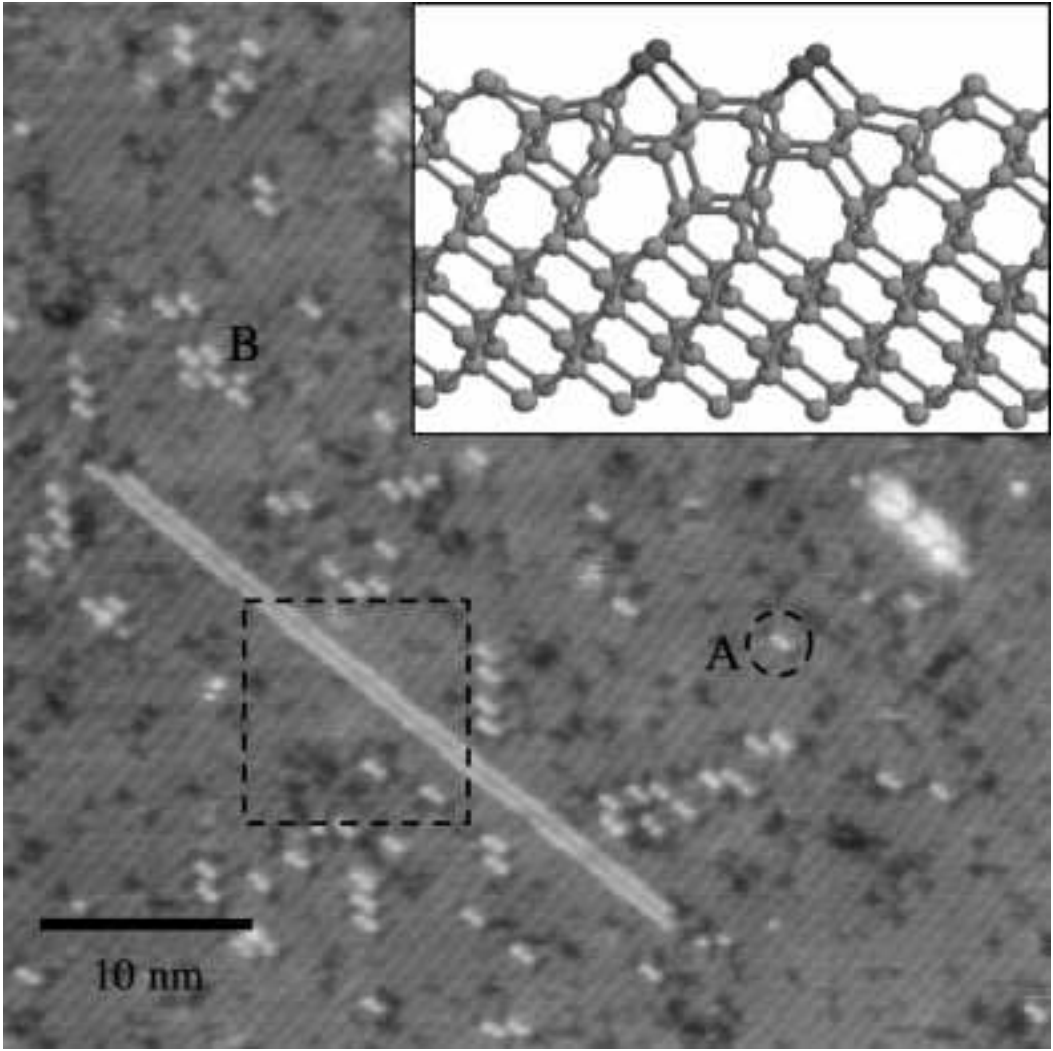} 
\caption{A 50 nm $\times$ 50 nm STM image of a small section of the
Si(001) surface containing a short nanoline, and numerous small
double-dot features (A). Note that some of the double-dot features
have formed small clusters in a diagonal or c$(4\times4)$ arrangement
(B).  The dotted box marks the area shown in
Fig.~\protect\ref{fig:varbias}. Inset: The Haiku structure for the Bi
nanoline\protect\cite{Owen2002b}. A pair of subsurface 7-membered
rings of Si form the basis of this unusual
reconstruction.\label{fig:bigimage}}
\end{figure}

Recently, we used a combination of atomistic structure calculations
and experiments to identify the structure of the
nanoline\cite{Owen2002b,Bowler2002} (which we have called the Haiku
structure).  An example of the Bi nanoline on a clean Si(001) surface
is shown in Fig.~\ref{fig:bigimage}, with the Haiku structure shown in
the inset. It has a complex reconstruction in the silicon substrate
reaching down five layers below the surface, though the surrounding
surface is not reconstructed.  The depth of the reconstruction
suggests that the nanoline may be coupled strongly to its
surroundings.  Knowledge of the nanoline structure allows us for the
first time to make detailed calculations of the electronic structure
of the nanoline, which bear out the observations made in STM.  The
calculations also provide an explanation for the additional features
that we have observed in recent atomic-resolution variable-bias STM of
the Bi nanoline on the clean Si(001) surface at room temperature.

In this letter, we present high resolution, variable bias images of a
Bi nanoline, and DFT calculations which explore the electronic
structure of the Bi nanoline.  In the next section, we discuss details
of the experimental and theoretical techniques used. We then present
the atomic-resolution images of the Bi-exposed Si(001) surface and
interpret them in light of the various calculations.

\section{\label{sec:details}Experimental and Theoretical Methods} 
The Si(001) substrate was cleaned using a standard
process\cite{Miki1998} before being transferred into vacuum. The Si
surface was prepared by flashing repeatedly to 1100$^\circ$C for a few
seconds, until there was only a small pressure rise. The clean surface
was checked with STM before Bi deposition began. Bi was evaporated
from an effusion cell, a typical dose being Bi at 470$^{\circ}$C for
10 mins. STM images were taken at the deposition temperature between
570-600$^\circ$C, and at room temperature, using a JEOL 4500 XT UHV
STM. The standard recipe for the formation of the Bi nanolines is to
anneal the Si(001) surface close to 600$^{\circ}$C, under a Bi flux of
around 1 ML/min.\cite{Miki1999a}. In this case, the surface was
annealed at a slightly lower temperature, around 570$^{\circ}$C, and
quenched to 300$^{\circ}$C as soon as the presence of nanolines had
been confirmed. In this way, we hoped to capture Bi nanolines while
still growing, and hence make some observations which would lead to a
likely nucleation mechanism. To minimise surface contamination, the
sample was held at 300$^{\circ}$C while the Bi cell was cooled and the
chamber pressure returned to its base level, before the sample was
cooled to room temperature.

The density functional theory (DFT) calculations were performed in the
Generalised Gradient Approximation (GGA)\cite{Wang1991,Perdew1992}
using the VASP code\cite{Kresse1996}, with ultrasoft pseudopotentials,
a plane wave cutoff of 150\,eV (sufficient for energy difference
convergence) and a Monkhurst-Pack \textbf{k}-point mesh with
4$\times$4$\times$1 points.  The unit cell used contained a single Bi
nanoline, and had ten layers of Si, with twenty atoms in each layer
(forming a single dimer row ten dimers long with the p($2 \times 2$)
reconstruction) with the bottom two layers constrained to remain fixed
and dangling bonds terminated in hydrogen. This unit cell was of
sufficient size to for convergence of energy with cell depth, and long
enough that the end Si dimers were representative of the clean,
undistorted surface.

\section{\label{sec:results}Electronic structure of Bi nanoline} 

Figure~\ref{fig:bigimage} shows a typical image from our STM
experiments.  Note that the Si(001) surface is clean, and that the
image was taken at room temperature (in contrast to all previous
experiments, where either the substrate was hydrogenated or the
temperature was $\sim570^{\circ}$C).  There are two key features to
note: first, the Bi nanoline, which is the double line extending
diagonally across the image from top left to bottom right; second, the
features (marked `A' in Fig.\ref{fig:bigimage}) which appear as a
double dot, and which are scattered across the surface; note that
these features tend to align diagonally with each other, even forming
a local c$(4\times4)$ pattern in places (marked `B').  In this paper,
we will concentrate on the electronic structure of the nanoline. We
are actively investigating a possible structure for the double-dot
features, which may be a precursor to the nanoline, and we shall
present our results in future work.

\begin{figure} 
\includegraphics[width=\textwidth]{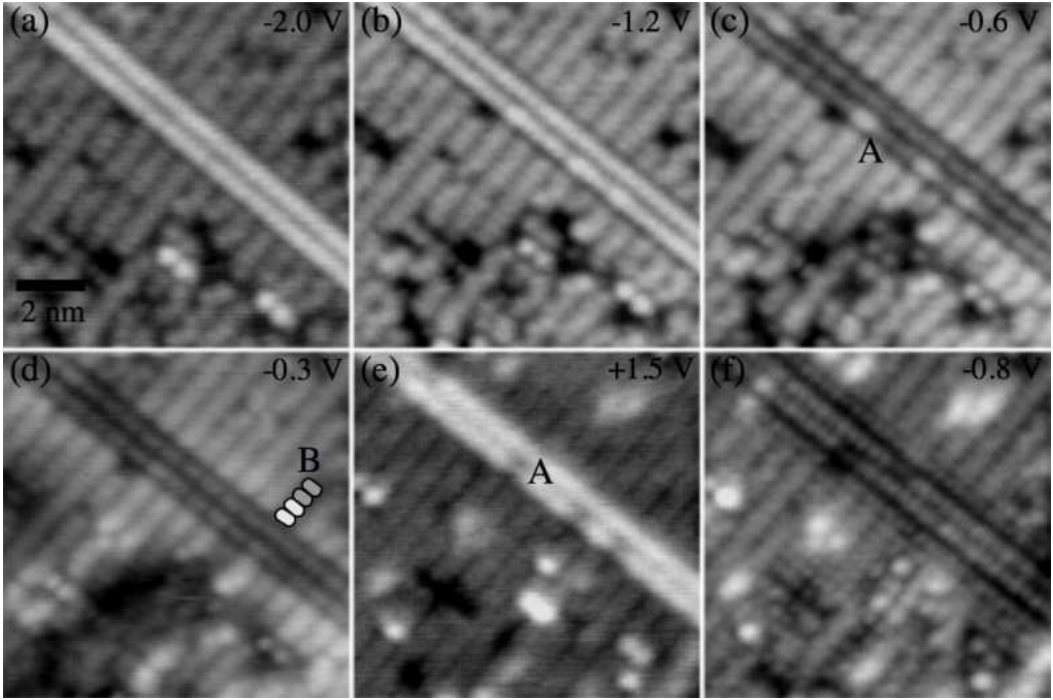} 
\caption{10\,nm $\times$ 10\,nm insets from the area shown in the
dotted black box in Fig.~\protect\ref{fig:bigimage}. The sample bias
voltages used are -2.0\,V, -1.2\,V, -0.6\,V, -0.3\,V, +1.5\,V and
-0.8\,V, in (a-f) respectively. As the sample bias is reduced, between
(a) and (d), the nanoline changes contrast from light to dark relative
to the surrounding Si(001). Over this range, some of the dimers in the
nanoline (marked A in (c) and (e)) exhibit a different voltage
contrast. At very low biases, around -0.3\,V, an enhancement of the
dimers around the nanoline, similar to that seen around a missing
dimer defect in clean Si(001)\protect\cite{Owen1995}, is seen.  This
is visible (and marked schematically as B) in (d). In (f), the
resolution is sufficient to see that the corrugation of the Si dimers
closest to the nanoline is increased, suggesting a greater separation,
and hence tensile strain.\label{fig:varbias}}
\end{figure}

In order to probe more closely the physical and electronic structure
of the nanoline, a series of atomic-resolution variable bias STM
images (in the area shown by the dotted box in
Fig.~\ref{fig:bigimage}) were taken and are shown in
Fig.~\ref{fig:varbias}.  The STM bias voltages are given in the figure
caption. Filled-states images range from -2.0\,V down to -0.3\,V. An
empty-states image at a bias voltage of +1.5\,V is also presented.  As
has been previously reported\cite{Miki1999a,Miki1999b}, the Bi
nanoline appears bright at large bias voltages of either sign, but as
the bias voltage is reduced, the contrast of the nanoline reduces so
that at around -0.8\,V, the relative contrast of the nanoline and the
Si substrate is the same, and at lower bias voltages, the nanoline
becomes dark relative to the clean Si surface. By contrast, images of
the Bi nanoline on the H:Si(001) surface\cite{Owen2002a} always show
the nanoline as a bright feature.

At very low biases, as in Fig.~\ref{fig:varbias}(d), substrate dimers
either side of the nanoline become enhanced, i.e. appear to be
brighter than the rest of the Si dimers, out to a distance of at least
two dimers.  (This is shown schematically in (d), by the lighter and
darker rectangles.)  This phenomenon of enhancement of neighbouring
dimers at low bias voltages has been seen previously for single
missing dimer defects (1DV) on Si(001)\cite{Owen1995}. In that case,
the enhancement was explained by the distortion of these dimers away
from their normal structure by the strain field of the 1DV, resulting
in a local change of the electronic structure, and the top-most
occupied states being moved higher in energy. The tensile strain
around the Bi nanoline is expected to be the cause of the enhancement
seen here; this enhancement is modelled below.  In
Fig.~\ref{fig:varbias}(f) (which shows extremely high resolution),
tensile strain near the Bi nanoline is seen directly: the darkening
between the first and second Si dimers is more pronounced than between
the second and third Si dimers, indicating an increase in separation
between these dimers.  In our DFT modelling, the dimer-dimer spacing
close to the nanoline is increased by 3\%, a significant strain.

Another feature visible in the images is a change of contrast between
Bi dimers in the nanoline.  While at elevated temperatures the Bi
nanoline always appears uniform, at room temperature, higher
resolutions may be achieved, and some subtle details resolved. Some
dimers in the Bi nanoline in Fig.~\ref{fig:varbias}(a) appear
different to the rest of the line (they seem to have a dark boundary
around them); these separated dimers correspond to brighter dimers on
the otherwise dark nanoline in the lower bias images (especially
Fig~\ref{fig:varbias}(c)) and darker patches (marked as `A') in the
empty states image shown in Fig.~\ref{fig:varbias}(e).  The reason for
these features have not been determined. However, they are not seen in
elevated-temperature images of the clean Bi nanoline, even at the same
bias voltages, while similar dark patches were observed during
experiments with adsorbed hydrogen and oxygen\cite{Owen2002a}. It is
therefore quite possible that they are the result of contamination by
ambient water or hydrogen while cooling down to room temperature.

\begin{figure} 
\includegraphics[width=\textwidth]{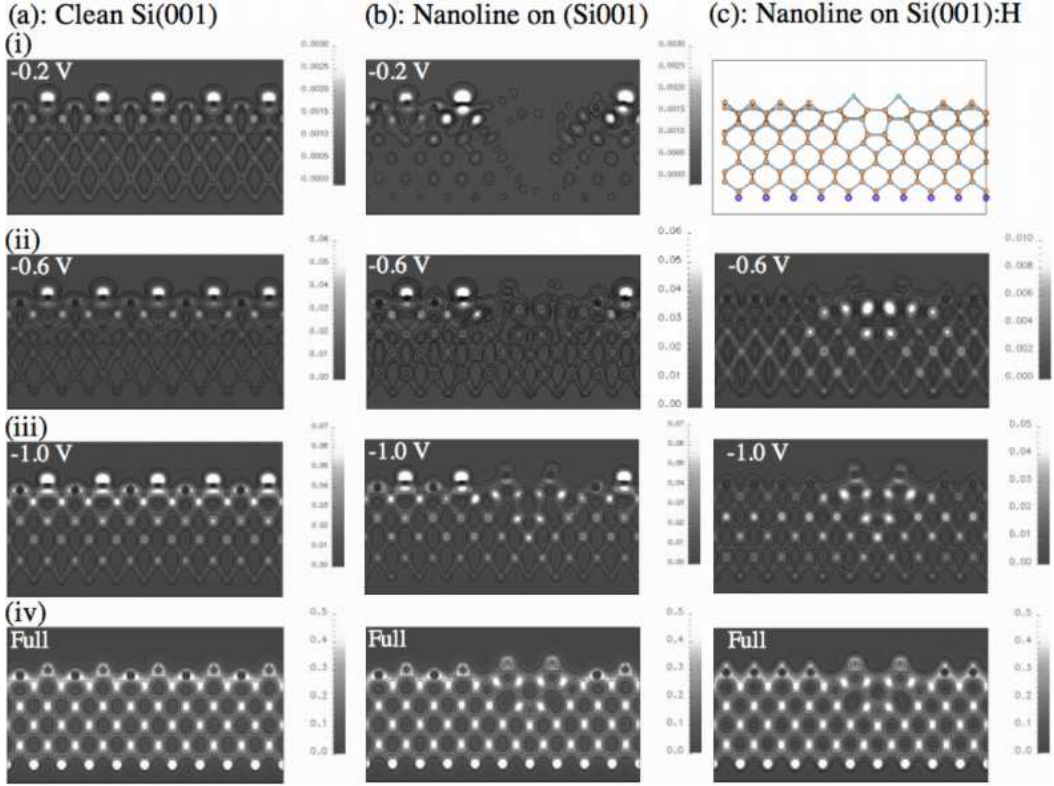} 
\caption{Contour plots of charge density for the states (i) 0.2\,eV,
(ii) 0.6\,eV and (iii) 1.0\,eV below E$_F$, and (iv) the total charge
density for: (a) the clean Si(001) surface (first column); (b) the
Haiku structure on the clean surface (second column), and (c) the
hydrogenated surface (third column).  The 0.2\,eV image for the
hydrogenated surface has been replaced with a ball-and-stick model of
the Haiku structure, as there are no states within 0.2\,eV of E$_F$.
See text for a full discussion.\label{fig:LDOS}}
\end{figure} 

We have calculated the electronic structure of the Haiku model for the
Bi nanoline using DFT, as described above.  In order to understand the
voltage contrast seen in STM, we have projected out the charge density
associated with various states within certain energies of the Fermi
level.  We show the partial charge densities associated with all
states within 0.2\,eV, 0.6\,eV and 1.0\,eV of the Fermi energy in
Figure~\ref{fig:LDOS} along with the complete charge density, for
three systems: (a) the clean Si(001) surface; (b) the Haiku model of
the Bi nanoline; and (c) the Haiku model with a hydrogenated Si(001)
surface.  For the last, the first image is missing as the first states
occur 0.55\,eV below the Fermi level (a side-on view of the Haiku
structure is shown in its place).  The partial densities are shown in
a plane passing through a Si dimer atom.  As the results are for a
statically buckled Si(001) surface, alternate dimers are ``up'' and
``down'' (except on the hydrogenated surface, where the buckling does
not occur). Only the up atoms appear bright, resulting in an
appearance of alternate missing dimers; the difference is most clearly
seen by comparison of the electron charge densities with the
structural model of the Haiku shown in the top right-hand corner of
Fig.~\ref{fig:LDOS}.

The partial densities are important for the comparison with
experiment, since the states close to the Fermi level will contribute
strongly to the STM images at low biases. Localisation of a state near
the Fermi level in the area around particular atoms or bonds (in
comparison to bulk or clean Si) is indicative of a local strain,
because strain forces structures away from their equilibrium state,
and hence minimum energy configuration. We first identified this
connection between strain and STM contrast (or states moved higher in
energy) in the case of the single missing dimer defect on
Si(001)\cite{Owen1995}, and we will use it here to elucidate the
substrate distortions around the Haiku structure.  In the case of the
clean Si(001) surface (first column of Fig.~\ref{fig:LDOS}), the
states near the Fermi level (a,i) are mainly localised on the surface
dimers, and the density fades away beneath the surface layer. This is
as expected, since the dimerisation of the Si(001) surface is
associated both with dangling bonds (at the surface) and with strain
(sub-surface). By contrast, the states near the Fermi level for the Bi
nanoline on clean Si(001) (b,i) are reduced in density below the Si
dimers, and are entirely excluded from the Haiku structure and
substructure. Instead, the two Si dimers immediately to the side of
the Haiku show a greatly enhanced charge density in the states within
0.2\,eV of the Fermi level, both by comparison with their neighbours
(the left-most dimers in (b,i)) and with the clean Si(001) surface
(a,i), which is seen in STM as the brightening of the dimers either
side of the Haiku at very low biases (Fig.~\ref{fig:varbias}(d) in
particular).  The localisation of these low-energy states indicates
that the substructure of the Haiku, far from being a strained
structure is \textit{more} relaxed than the layers beneath a clean
Si(001) surface. Furthermore, it demonstrates that the strain induced
by the Bi nanoline is concentrated on the Si dimers either side of the
Haiku structure, and on the second-to-third layer bonds beneath the
dimers.

The electronic structure of the Bi nanoline with a H-terminated
Si(001) surface is shown in Fig.~\ref{fig:LDOS}(c).  The major effect
is the removal of the surface states associated with the Si dimer
$\pi$-bonds, and the removal of the buckling on the surface.  The Si
atoms immediately beneath the Bi nanoline now show an
\textit{increased} charge density relative to the neighbouring Si
atoms, which is clearly visible in the 0.6\,eV and 1.0\,eV images
(c,ii \& iii), indicating that the region below the Bi nanoline is
more strained than bulk Si. Furthermore, enhancement can be seen in
layers as deep as the sixth layer, indicating the deep strain caused
by the presence of the Bi nanoline. This difference in the electronic
states suggests that the contrast seen with the clean surface is due
to the interaction of the buckled dimers and the nanoline (and in
particular the ``up'' atom of a buckled dimer) which is removed with a
hydrogenated surface.

The lack of any states on the Bi near the Fermi level is clearly the
cause of the darkening of the nanoline relative to the Si(001) surface
seen in STM at low voltages.  Indeed, the first states seen on the
nanoline lie over 0.5\,eV away from the Fermi level (which agrees
qualitatively with the bias voltage of 0.8\,V where parity of
appearance between the nanoline and the substrate occurs).  In the
full charge densities(Fig.~\ref{fig:LDOS}(iv)), the density of states
associated with the Bi dimers, combined with their higher physical
height, explains their relative brightness in STM at higher bias
voltages. In the case of the hydrogenated surface, the Bi dimers have
a similar charge density as in the clean surface, but in this case,
the Si $\pi$-bonds have been eliminated, so the Bi dimer remains
bright in STM at all biases\cite{Owen2002a}.  The large density
associated with the nanoline in the 0.6\,eV(c,ii) and 1.0\,eV(c,iii)
images underline the mechanism behind the brightening of the Bi
nanoline.

\section{\label{sec:conc}Conclusions} 
We have presented high resolution room temperature STM images of Bi
nanolines on a clean Si(001) surface for the first time.  These images
allowed us to probe the electronic structure of the nanoline and the
surrounding Si(001) surface, and, by extrapolation, the localised
strain associated with the Haiku. We found that the two Si dimers
neighbouring the nanoline show enhancement at low bias voltages, and
we confirmed that the nanoline is darker than the surrounding surface
for biases of less than 0.8\,V.  We have also presented first
principles electronic structure calculations of the Haiku structure
with both a clean and a hydrogenated Si(001) surface, and we have
explained the voltage contrast of the nanolines using these
calculations.

The localisation of the states near the Fermi level seen in the
projected charge densities suggests that the Si atoms immediately
surrounding the Haiku structure are strained (in part at least due to
the interaction with the buckled Si dimers), while the Si substructure
immediately beneath the Haiku itself is somewhat relaxed (though the
hydrogenated surface plots suggest that it is slightly strained
relative to perfect bulk Si). This goes some way to explaining the
stability of the subsurface 5-7-5 ring structures, which serve as a
highly effective relief mechanism for the epitaxial stress exerted by
adsorbed Bi.  The absence of states close to the Fermi level suggests
that the nanolines are likely to block surface conduction
perpendicular to the nanoline, and are not likely to act as a
nanowire.

The Bi nanolines are clearly not suitable for conduction on their own.
However, as we have shown before\cite{Owen2002a}, hydrogen will adsorb
preferentially on the Si(001) surface and not on the Bi, giving a
natural, automatic masking technique.  This should allow us to adsorb
metals on the nanolines, creating exceptionally high quality
nanowires.  This research is under way, and will be presented in
future work.

\ack 
DRB thanks the Royal Society for funding through a University Research
Fellowship. Calculations were performed at the HiPerSPACE Centre at
UCL (JREI grant JR98UCGI). JHGO was supported by the Japanese Science
and Technology Agency (JST) as an STA Fellow. This study was performed
through Special Coordination Funds of the Ministry of Education,
Culture, Sports, Science and Technology of the Japanese Government
(Research Project on active atom-wire interconnects). We would like to
thank Bill McMahon for bringing to our attention many of the examples
of 5-7-5 structures, and for sharing unpublished STM data.

\bibliography{SSLBiElecStruc} 

\end{document}